\documentstyle[a4wide,epsfig,12pt]{article}
\begin{document}
\begin{flushright}
\begin{obeylines}
NORDITA - 96/49 N,P
hep-ph/9607460
\end{obeylines}
\end{flushright}
\vskip 1cm
\thispagestyle{empty}
\centerline{\Large Neural Network analysis for $ \gamma \gamma \to 
\pi^+ \pi^- \pi^0$ at Daphne
\footnote{This research is partly
supported by EU under contract number CHRX-CT92-0004 and by the
Comissionat per Universitats i Recerca de la Generalitat de Catalunya.}
}
\vskip 1.2cm
\centerline{Ll. Ametller$^a$, Ll. Garrido$^{b,c}$ and 
P. Talavera$^d$ \footnote{Supported by EU under contract number 
ERB 4001GT952585.} }

\vskip 1cm
\centerline{$^a$Departament de F\'\i sica i Enginyeria Nuclear}
\centerline{Universitat Polit\`ecnica de Catalunya, E-08034 Barcelona, Spain}
\vskip 0.3cm
\centerline{$^b$Departament Estructura i Constituents Mat\`eria} 
\centerline{Universitat de Barcelona, E-08028 Barcelona, Spain}
\vskip 0.3cm
\centerline{$^c$Institut de F{\'\i}sica d'Altes Energies}
\centerline{Universitat Aut\`onoma
de Barcelona, E-08193 Bellaterra (Barcelona), Spain}
\vskip 0.3cm
\centerline{$^d$ NORDITA, Blegdamsvej 17, DK-2100, Copenhagen \O, Denmark}
\vskip 3cm
\begin{abstract}
We consider the possibility of using neural networks in experimental
data analysis in Daphne. We analyze the process $\gamma\gamma\to \pi^+
 \pi^-  \pi^0$ and
its backgrounds using neural networks and we compare their performances 
with traditional methods of applying cuts on several kinematical variables. 
We find that the neural networks are more efficient and can be 
of great help for processes with small number of produced events.
\end{abstract}
\newpage

\section{Introduction}

The Daphne $\Phi$--factory should start to operate very soon in Frascati.
Its main goals are the study of $\Phi$ decays and related processes, mainly
studies of CP violation in kaon decays, $\pi$--$\pi$ phase shifts, $\eta$
decays, etc. \cite{DAPHNEBOOK}
Being an $e^+$ $e^-$ collider, the machine is also suited to
study $\gamma\gamma$ physics. In this context, the golden plate process
at Daphne is $\gamma \gamma\to  \pi^0 \pi^0$. The main reason is 
that the present experimental situation is not well established, at least 
at the region near threshold, where good theoretical predictions exist
in the context of Chiral Perturbation Theory (ChPT). Moreover, this
theoretical predictions start at the one-loop level, and thus this process
is a clear test of the effective quantum field theory character of ChPT.
In a similar way, other interesting processes have been proposed recently.
In particular, the processes $\gamma\gamma \to 3 \pi$ are also interesting
because one--loop predictions dominate over tree-level ones
\cite{TABBC}. They differ
however from $\gamma \gamma\to	\pi^0 \pi^0$ in several aspects: i) They are
anomalous processes, ii) they are not exclusive
test of chiral loops since they get contributions from counterterms and iii)
their cross sections are much smaller, thus more difficult to measure 
experimentally. The best way consists in
tagging the electron and positron, but at expenses of reducing significantly
the number of events due to small tagging efficiencies \cite{zallo}. 
It is therefore
convenient to dispose of alternative methods with large efficiency and without
lepton tagging whenever possible. We suggest that neural networks (NN's)
could be used in experimental analysis for such a purpose. We have trained a
NN with $\gamma \gamma\to \pi^+ \pi^-\pi^0$ (signal) and have considered
the three main sources of background. Our analysis avoids tagging (thus we are not
penalized by small tagging efficiencies) and obtains results which are 
better than traditional methods based in applying cuts over a set of
kinematical variables. 

This work is organized as follows. In section 2 we
describe briefly ChPT and its predictions for $\gamma\gamma\to 3\pi$ at
Daphne. Section 3 gives a short description of NN's. In Section 4 we
describe the generation of data for the signal and the analyzed backgrounds 
and introduce the set of
kinematical variables which are used as the NN inputs. The performance of
the NN is compared with the usual methods of analysis in Section 5. Section
6 is devoted to the conclusions.

\section{Chiral Perturbation Theory for $\gamma\gamma\to 3\pi$}

ChPT is an effective formulation of QCD at 
low energy in terms of pseudoscalar mesons as fundamental fields\cite{CHPT}. 
It is inspired from QCD enforcing its symmetry properties. Indeed, the QCD
Lagrangian  --in terms of quarks and gluons-- possesses a Chiral 
$SU(3)_L\times SU(3)_R$ symmetry, for massless
quarks. However, when considering the quark mass terms, these break the
chiral symmetry. In the effective low energy version of QCD, one replaces the
fundamental quark and gluon fields by the pseudoscalar mesons, imposing 
$SU(3)_L\times SU(3)_R$ symmetry, only broken by terms proportional
to the quark masses, which can be related to the pseudoscalar
meson masses. The ChPT Lagrangian can be written as an expansion in momenta
and masses 
and treated perturbatively. The first order ChPT predictions 
are essentially equivalent
to Current Algebra. However, they get corrections from higher order
terms, playing loops a particular role. They are essential in order to
incorporate the correct analyticity, unitarity and 
crossing symmetry properties of the physical amplitudes. 
Moreover, loops give --in general-- divergences
which can only be absorbed by tree-level counterterms 
present in the higher order Lagrangian. The number of needed 
counterterms depends on the order of the momentum expansion.
This is a consequence of the non renormalizability of the theory (in the 
classical sense that all divergences can be absorbed in a fixed, finite, 
number of terms).
In spite of that, the theory can still give predictions provided
one can fix the values of the counterterms through related processes. 
When this is not possible, one has to rely on some phenomenological models 
to estimate those counterterms, but at expenses of introducing model
dependence in the game. 
Electroweak interactions can be introduced in a systematic and
selfconsistent way.

One distinguishes two sectors in ChPT. The normal, even intrinsic parity
sector, treats processes as, for instance,
$\pi \pi \to \pi\pi$, $\eta\to 3\pi$, or
$\gamma\gamma\to \pi\pi$. The anomalous, odd intrinsic parity sector, 
accounts for processes 
as $\pi^0\to\gamma\gamma$, $\eta\to\pi\pi\gamma$, and 
$\gamma\gamma\to 3\pi$. From the former sector, 
the $\gamma\gamma\to \pi^0\pi^0$ process plays an
important role. This is because there is no tree level contribution 
\cite{BIJCOR}. The 
first non vanishing contribution starts at $O(p^4)$ --in the momentum
expansion-- and is 
entirely given by one loop diagrams, with no contribution from the 
$O(p^4)$ tree level Lagrangian. This makes the process very interesting, 
since it 
tests the loop predictions of the theory. In a similar way, the anomalous
processes $\gamma\gamma\to 3\pi$ receive contributions from $O(p^6)$ which
dominate over the non vanishing tree level ones and test the 
loop predictions in the anomalous sector, although in a less severe way,
since there are two types of $O(p^6)$ contributions, loops and counterterms.

In Refs.\cite{TABBC,ABBCT} the amplitudes for $\gamma\gamma \to
\pi^+\pi^-\pi^0$ and $\gamma\gamma \to 3\pi^0$ have been obtained with
the corresponding predictions for the expected number of events at Daphne.  
One expects around $180$ ($23$) events per year for the first (second) 
process. These are quite moderate number of events, and require good
strategies for its eventual experimental detection. 
We suggest that using
NN's can be a good possibility to perform an efficient analysis.
We restrict our analysis to the first, charged channel, which looks a priori
more promising that the neutral one.

\section{Neural Networks}

Neural Networks (NN's) are useful tools for pattern
recognition. In high energy physics, they have been
used or proposed as good
candidates for tasks of signal versus background classification. Some
examples are the Higgs searches \cite{ChSt}, 
b and $\tau$
analysis \cite{L3}, quark and gluon jets analysis \cite{Jets},
determination of Z to heavy quarks branching ratios \cite{Z}, bottom-jet
recognition \cite{Odorico} and top--quark search in $p\bar p$
colliders \cite{AGT,AGT2}.
Recently, NN's have been used for experimental
top quark searches at the Tevatron\cite{Push}.

We have considered layered feed--forward NN's with topologies 
$N_i \times N_{h_1} \times N_{h_2} \times N_o$,
where $N_i$ ($N_o$) are the number of input (output) neurons  and
$N_{h_1}, N_{h_2}$ are the neurons in two hidden layers.
 
The input of neuron $i$ in layer $l$ is given by,
\begin{equation}
\label{neusum}
I^{l}_{i}  =in^{(e)}_{i}, \qquad l=1  \;\; ; \qquad\qquad
I^{l}_{i}  =  \sum _{j} w_{ij}^{l} S_{j}^{l-1}+ B_{i}^{l},  \qquad l=2,4\;\; ,
\end{equation}
where $in^{(e)}_{i}$ is the set of kinematical variables describing
a physical  event $e$,
the sum is extended over the neurons of the preceding layer
$(l-1)$, $S_{j}^{l-1}$ is the state of the neuron $j$,
$w_{ij}^{l}$ is the connection weight between the neuron
$j$ and the neuron $i$, and  $B_{i}^{l}$ is a bias input to neuron $i$.
The state of a neuron is a function of its input
$S_{j}^{l}=F(I^{l}_{j})$, where $F$ is the
neuron response function. In this study the ``sigmoid function'',
$F(I^{l}_{j})=1/(1+exp(-I^{l}_{j}))$, has been chosen. This function offers a more sensitive modeling of real data than a linear one.
 
Back--propagation was used as the  learning algorithm.
Its main objective is to minimize the quadratic output-error $E$,
 
\begin{equation}
\label{energy}
E=E(in^{(e)}_i,out^{(e)},w_{kl},B_k)=
\frac{1}{2}\sum_{e}(o^{(e)}-out^{(e)})^2 \ .
\end{equation}
This minimization is obtained
by adjusting the $w_{kl}$ and $B_n$ parameters, where
$o^{(e)}$ is
the state of the output neuron for event $e$,
$out^{(e)}$ is its desired state, and $e$
runs over the learning sample.
Taking the desired output as 1 for
signal events and 0 for background events, the network output
gives, after training, the conditional probability that
new test events presented to the network are of signal- or background-type
\cite{Lluis},
provided that the signal/background ratio used in the
learning phase corresponds to the real one.
 
Weights are updated for each event presented to the NN during the
learning phase.
Once the quadratic error $E$ reaches its minimum value,
they are kept fixed and used in the testing phase where the NN is
used as a signal--background classifier.
A frequent problem encountered
in NN training is over-learning. It takes place when the NN interprets
statistical fluctuations as real differences. In this study
over-learning is
avoided by checking the evolution of the error  on a
test sample, $E_t$, and stop learning when $E_t$ starts to increase,
even if the learning error  function $E$ still continues to decrease.

\section{Data generation for signal and backgrounds}

We take as signal in our analysis the process $\gamma\gamma\to\pi^+ \pi^-
\pi^0$ as predicted by ChPT at $O(p^6)$ \cite{ABBCT} in Daphne, running
at $e^+ e^-$ center of mass energies $\sqrt{s}=M_\Phi$.
We avoid tagging of the leptons for its analysis, in order to keep all
produced events. In so doing, we had to consider several types of
backgrounds.  We analyzed the following ones: 

\begin{itemize}
\item[B1)] \qquad $\gamma\gamma\to\eta\to\pi^+ \pi^-\pi^0$
\item[B2)] \qquad $e^+ e^-\to \omega,\Phi\to \pi^+ \pi^-\pi^0 \gamma$
\item[B3)] \qquad $e^+ e^- \to (\omega, \Phi) \gamma \to\pi^+ \pi^-\pi^0 \gamma$
\end{itemize}

The first background B1 is eta photoproduction. It has the same origin as the
signal and differs from it because the invariant mass of the three pion
system is strongly peaked around the eta mass. 
The background B2 consists in the decay of a real (virtual) $\Phi$ ($\omega$)
vector meson which decays into $\pi^+ \pi^-\pi^0 \gamma$. 
The background B3 accounts for virtual production of $\Phi$ or $\omega$ vector
mesons and one initial state bremsstrahlung photon.
In the last two processes, we demand the presence of one undetected photon.
This photon decreases the available energy of the three-pion system and 
eliminates the production and decay of a virtual $\omega$ into $3 \pi$ 
as potential background.
The photon escapes detection mainly going through the beam pipe. 

As it has been previously mentioned, the signal has been predicted in the 
context of ChPT. The first background has been estimated in the same
context, but using a constant matrix element  evaluated at the center of the
Dalitz plot. This is a good approximation for our purpose of estimating the
range of energies where this background is important. The backgrounds B2 and
B3 have been computed using vector meson dominance. 

We have generated Monte Carlo events for the signal and the backgrounds, 
satisfying the following generation cuts:

1) All pions are in the detector, which has almost full $4 \pi$ acceptance 
except for the beam pipe, which corresponds to a fraction of $2\%$ 
of the total solid angle.

2) The invariant mass of the three pion state is restricted to be in the range
$ 3 m_\pi \le m_{3\pi} \le 0.7$ GeV. The upper limit is conservatively taken
in such a way that the ChPT matrix elements used for the signal and the 
background B1, computed at $O(p^6)$, can be 
trusted. For larger invariant masses, one expects that higher order corrections
could be important and modify significantly the estimation of the signal.

3) The photons produced in backgrounds B2 and B3, escape detection through
the beam pipe. On the contrary, they should be easily detected since their 
energy is forced to be in the range $255$~MeV~$\le E_\gamma \le 341$~MeV
due to the above constraint imposed on $m_{3\pi}$.

4) As we are interested in a final $3\pi$ state not coming from $\eta$ 
production, we make an additional cut on $m_{3\pi}$. We demand
$ m_\eta + \Delta \le m_{3\pi} \le m_\eta - \Delta$. Taking $\Delta=
20$ MeV, the first background is practically eliminated \cite{ABBCT}, 
since it is only important around the $\eta$ mass region. 

The number of expected events per year passing the generation cuts for the 
Daphne integrated luminosity of $\int L dt=5\times 10^6$~nb$^{-1}$ 
are: 71, 0.07, 1714, 776
for the signal and each of the backgrounds, respectively. 
The background B1 can be safely discarded.

For the analysis and as inputs to the NN, we chose the following kinematical 
variables 

\begin{itemize}
\item 1) The $\pi^+$ transverse momentum,
\item 2) the $\pi^-$ transverse momentum,
\item 3) the $\pi^0$ transverse momentum,
\item 4) the three pion system transverse energy,
\item 5) the $\pi^+$ pseudorapidity,
\item 6) the $\pi^-$ pseudorapidity,
\item 7) the $\pi^0$ pseudorapidity,
\item 8) the three pion system sphericity in the three pion center of mass,
\item 9) the difference between $m_\rho$ and its best approximation
by the invariant masses of all possible pairs of pions. 
\end{itemize}

All the above variables are self explained, except the last one which was
chosen because background B3 is mediated by a $\rho$ exchange with its
subsequent decay into a pion pair. There is no need to say that one could have
considered other variables, as angular correlations among the final pions, 
for example, which bring additional information on the physics of the process
and could help in the task of signal versus background separation.

\section{Results}

Rather than using the expected number of events produced at Daphne, we 
generated bigger samples of $10000$ signal, B2 and B3 background events, 
passing the generation cuts. From each of those, $8000$ events were used to
train  a $(N_1=9) \times (N_{h_1}= 11) \times (N_{h_2}=5) \times (N_o=1)$   
NN --denoted by NN9 from now on-- to give output 1 for the signal and 0 for the
backgrounds. 
The rest of events were reserved for doing the NN test and the analysis 
in a classical way. 
(The B1 background is eliminated by the generation cut 4.)
The obtained results were rescaled to the 
expected number of events produced at Daphne per year.
In Fig. 1 we show the distribution of the test events that survive as a 
function of the NN9 output cut. (An event survives if its corresponding output
is larger than the chosen output cut.) The dot-dashed (solid) line 
corresponds to the signal (total background) events. 
One can deduce that the signal events are 
very peaked to output values very close to one, while the background events 
tend to concentrate at values close to zero.
It is clear that one can select subsamples richer on signal or background 
with suitable choices of NN output cuts. 
In our case, we are interested in improving the signal to background ratio,
thus we will accept events with outputs larger than a given output cut.
A good variable to parametrize the efficiency of the analysis is the 
statistical significance, defined as $S_s=N_s/\sqrt{N_b}$, being $N_s$ 
($N_b$) the number of signal (background) accepted events. 
The solid  line in Fig. 2 shows the statistical significance $S_s$
as a function of the NN9 output cut. The curve has 
been plotted for output cut values up to $0.95$, to avoid strong fluctuations
on its estimation due to lack of statistics.
For output cuts around $0.9$, the achievable  $S_s$ is around $60$, thus
indicating that the NN performs a very good job in the signal recognition
against the considered backgrounds.  

At this point, we would like to stress the benefits of using the NN over
more traditional methods of doing the experimental analysis. Indeed,
usually experimentalists perform several cuts on some kinematical variables
to isolate the regions where the signal differs most from the backgrounds.
This procedure, when one considers a large number of variables, is usually
done by means of linear cuts, isolating hypercubical regions in favor of the
signal. Its efficiency is known to be lower than the achievable by NN
techniques \cite{AGT}. One can wonder, however, how the NN results compare 
with smarter ways of applying cuts. In particular, it is clear that
one has to isolate regions of the parameter space with complicated geometry,
so linear cuts will have limited success in general. One has to consider
non linear cuts specially designed by previous inspection of the signal
and background. This is only feasible, in practice, for small number
of variables in the analysis.  
We performed an analysis in these terms using the three most 
significant variables. These were obtained from the nine original variables 
using the methods discussed in Ref. \cite{AGT2}, involving the weights
connecting the inputs with the first hidden layer.
They turned to be the pseudorapidities of the pions. The topology
of the signal and background events in the three pseudorapidities space
look qualitatively different. The signal tends to lie inside an ellipsoidal
surface centered at the origin, while the background events are 
preferably distributed into two separated regions, symmetrically located
respect to the origin.

We could isolate non linear regions with statistical significances up to 
$10$. Notice that this result must not be compared with the results of the
NN9, which were obtained using the full set of the original $9$ variables.
In order to do a fair comparison, we trained a smaller NN using the same three 
input variables, which we denote by NN3, with topology 
 $(N_1=3) \times (N_{h_1}= 5) \times (N_{h_2}=5) \times (N_o=1)$. 
The results of this NN3 net are also shown in the figures. The dotted (dashed)
line in Fig. 1 shows the accumulated number of signal (background) events as 
a function of the NN3 output cut. The reconstruction of the signal is fairly 
good, but the background is much worse respect to the NN9. This
translates into much smaller statistical significances, typically by a factor
of $6$, as it is shown by the dashed line of Fig. 2, where $S_s$ is plotted
as a function of the NN3 output cut.

Two comments are in order. First, notice that we were interested in reducing
the variables to three, to be able to design good sets of non 
linear cuts with the help of three dimensional distributions
of events. In case of keeping more variables, the statistical significances
would not be so drastically reduced.
Second, the reduced NN3, for output cuts larger than $0.85$, 
is at least as efficient as the best optimized non linear classical
cuts we could find. Moreover, there is a great advantage of the NN3 in front of
the non linear cuts: Whereas the latter have to be designed by visual
inspection and require dedicated work, the NN operates in a completely
automatic way, with comparable efficiency.
This is not surprising. It is due, in fact, to the highly non linear 
behaviour of the NN's, which allows them to select 
complicated regions of the parameter space in an automatic and painless way. 

Finally, NN's can be easily trained for any number of input variables.
On the contrary, non-linear classical analyses are strongly limited to 
small number
of variables. This makes NN's very useful tools for processes where high
efficiencies are needed.

\section{Conclusions}

We have considered the ability of NN's to perform experimental analysis 
for the process $\gamma\gamma\to \pi^+\pi^-\pi^0$ at Daphne. The ChPT 
prediction for the number of events produced is relatively small, thus
indicating that efficient methods for its detection and analysis can be
of great help. We have considered three types of backgrounds which
mimic the signal and we have avoided tagging of the initial leptons
which would imply a sensible reduction of event statistics.
Using a set of nine kinematical variables as inputs of a NN9,
we have obtained large statistical significances for a wide interval of
output cuts.

We have also studied the expected 
efficiencies for a smaller NN3, using the three pion pseudorapidities as 
inputs, and compared them with the efficiencies found by
using classical analyses in terms of non linear cuts for
the same variables.
We have found that the NN3 statistical
significances, obtained in an automatic and painless way, are 
at least as good as the best result we could find using non linear cuts 
chosen through accurate inspection
on the distribution of the signal and background events in the three variable
space.
This is due to the highly non linear behaviour of the NN, which 
isolates the phase space regions where the signal 
differs significantly from the background. 
However, the NN3 efficiency is
much smaller than the one obtained by NN9. It is therefore highly 
recommended to use large sets of kinematical variables for ensuring large
efficiencies. This represents no extra effort for the NN's and can be a 
great challenge for classical methods.
We finally stress that the usefulness of NN's is not restricted to
the signal analysed, but it can be shown to work similarly for any other 
process of interest.

\vskip 1cm
\centerline{\bf Acknowledgements}
\vskip .8cm
We thank A. Bramon and J. Bijnens for useful discussions.

\vskip 1.5cm
\listoffigures
\begin{figure}
\epsfig{file=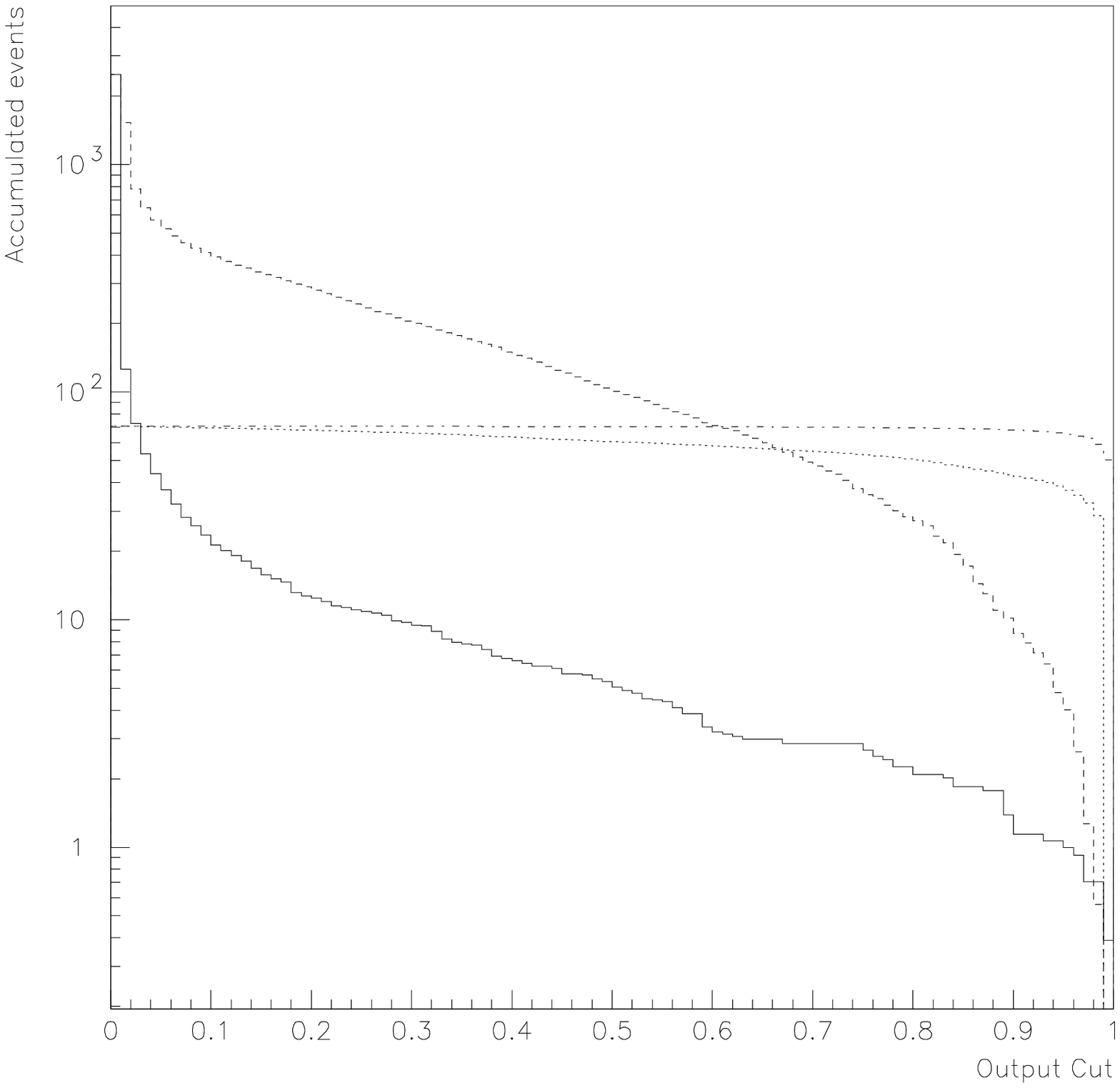,width=14cm,height=16cm,angle=0}
\caption[Accumulated number of events with output larger than a given 
output cut.]
{Accumulated number of events with output larger than a given output cut. 
Solid (dashed) and dot-dashed (dotted) lines correspond to the background 
and signal results, respectively, for NN9 (NN3).}
\end{figure}
\begin{figure}
\epsfig{file=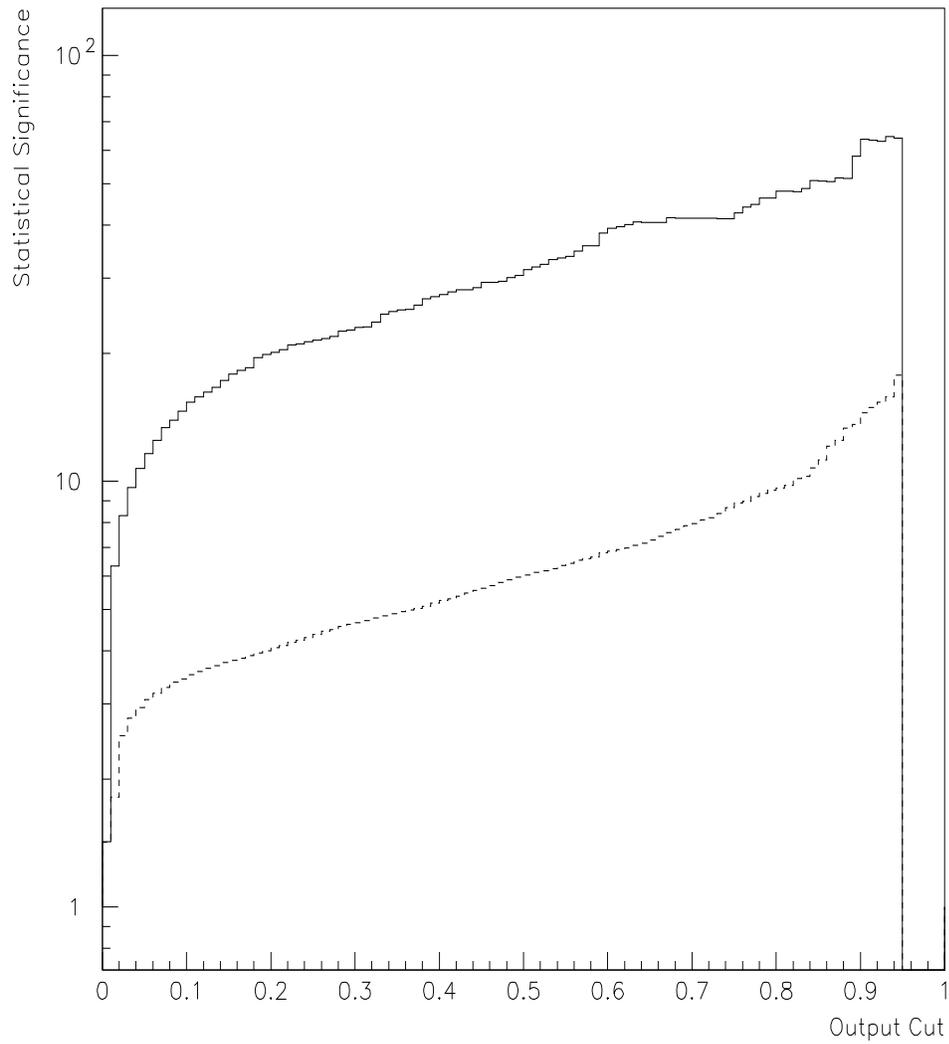,width=14cm,height=16cm,angle=0}
\caption[Statistical significance as a function of the NN output cut.]
{Statistical significance as a function of the NN output cut. Solid (dashed)
line corresponds to the NN9  (NN3) net.}
\end{figure}
\end{document}